\documentclass[showkeys,superscriptaddress]{revtex4-2}
\usepackage{tabularray}
\usepackage{amsmath,graphicx,array,float,amsthm,amssymb}
\usepackage{tikz,color,fullpage,epsf,caption,subcaption,multirow}
\usepackage{amsmath,graphicx,verbatimbox,array,float}
\usepackage{indentfirst,orcidlink,hyperref}
\usepackage{braket}
\usepackage{diagbox}
\usepackage{amssymb}
\usepackage{bbold}

\newcommand{\figref}[1]{Fig.~\ref{#1}}

\renewcommand{\eqref}[1]{Eq.~(\ref{#1})}

\theoremstyle{definition}
\newtheorem{definition}{Definition}

\renewcommand{\eqref}[1]{Eq.~(\ref{#1})}

\begin{document}
\title{Comment on \\
Quantum illumination using polarization-entangled photon pairs for enhanced object detection (Opt. Express 32, 40150-40164, 2024)}

\author{Artur Czerwinski \orcidlink{0000-0003-0625-8339}}\email{aczerwin@umk.pl}
\author{Jakub J. Borkowski \orcidlink{0009-0007-1526-9394}}
\affiliation{Institute of Physics, Faculty of Physics, Astronomy and Informatics, Nicolaus Copernicus University in Torun, ul. Grudziadzka 5, 87-100 Torun, Poland}

\begin{abstract}
The paper by K. Sengupta et al. (Opt. Express 32, 40150-40164, 2024) explores quantum illumination using polarization-entangled photon pairs for object detection in noisy environments. In this comment, we highlight fundamental flaws in the mathematical model used to describe photon loss. We argue that the treatment of photon loss and its effects on quantum entanglement is incorrect. We demonstrate that the conclusions of Sengupta et al., particularly the detection of low-reflectivity objects using quantum correlations, are unsubstantiated, as the assumed resilience of polarization entanglement to photon loss contradicts established principles of quantum information theory. We present a more rigorous framework for describing the effects of photon loss on both polarization-encoded and photon-number quantum states. Additionally, we critique the approach used in the OE article to model photon loss in free-space optical (FSO) transmission, noting that it is based on a fiber-optic model that was adopted with insufficient attribution from an earlier publication. We propose several improvements to enhance the modeling of FSO photon loss.
\end{abstract}


\maketitle

\section{Introduction}
The paper by K. Sengupta et al. (Opt. Express 32, 40150-40164, 2024) \cite{Sengupta2024} experimentally demonstrates quantum illumination using polarization-entangled photon pairs to detect low-reflectivity objects in noisy environments. By utilizing the Clauser-Horne-Shimony-Holt (CHSH) inequality and its normalized value as an indicator of non-classical correlations, the authors claim to detect objects with reflectivity as low as $0.05$. The study explores the robustness of these correlation measures against photon loss, noise, and depolarization, and introduces residual quantum correlation for enhanced detection under reduced quantum conditions. The feasibility of this approach is further analyzed by modeling photon attenuation and assessing performance in atmospheric conditions.

In this paper, we address several inaccuracies in Ref. \cite{Sengupta2024}, focusing particularly on Sec. 3.3, "Attenuation and range estimation," where the authors analyze CHSH values in the context of free-space optical (FSO) transmission. To achieve this goal, they apply a mathematical model that treats attenuation as a non-unitary decoherence process affecting the Fock state representing the number of photons. This formalism was introduced in Ref.~\cite{Czerwinski2022}, where it was applied to examine the impact of attenuation on quantum state tomography (QST) and entanglement quantification. For clarity, henceforth we will refer to Ref. \cite{Sengupta2024} as "the OE article" and to Ref. \cite{Czerwinski2022} as "the PRA article."

This commentary is organized as follows. First, in Sec. \ref{section0}, we enumerate several critical flaws in the OE article, focusing on key assumptions and the mathematical formalism used in the referenced paper. In Sec. \ref{section2}, we present an accurate mathematical model for describing the effects of photon loss in entangled states, which we believe should have been adopted instead of the approach taken in the OE article. Next, in Sec. \ref{section1}, we outline and discuss the overlaps between the OE and PRA articles, noting that Sec. 3.3 of the OE article lacks proper citation of the PRA article, despite significant overlaps suggesting reliance on its contributions. Then, in Sec. \ref{section3}, we argue that the attenuation model used in the OE article for FSO transmission is oversimplified and inadequate for a comprehensive study of photon loss. We propose a more detailed model and compare our results with those from the OE article. Lastly, we summarize our findings and offer additional comments in Sec. \ref{finalsec}.

\section{Critical Flaws in the OE Article}\label{section0}

We begin by enumerating several critical issues that affect the overall quality of the OE article. These comments primarily concern the treatment of the Kraus operators and the assumptions regarding quantum entanglement subject to a lossy channel.

\subsection{Kraus operators for a lossy channel}

In Eq. (6) of the OE article, the authors correctly identify the Kraus operators that describe a lossy channel in the Fock space. Let $\ket{0}$ and $\ket{1}$ represent the Fock states corresponding to zero and one photon, respectively. According to the OE article, photon loss may only occur for the signal photon, which leads to the following forms for the Kraus operators (for consistency we retain the same equation number as in the OE article):
\begin{equation}\label{eq1}
    K_0^s = \sqrt{1 - \eta } \ket{0}_s \!\bra{1}_s \hspace{0.75cm}\text{and}\hspace{0.75cm}  K_1^s = \ket{0}_s \!\bra{0}_s + \sqrt{ \eta } \ket{1}_s \!\bra{1}_s \tag{6},
\end{equation}
where $\eta$ denotes the transmittance of the channel (referred to as {\it reflectivity} in the OE article). Then, for convenience, let us denote the lossy channel by $\Lambda_{\eta}$.

The operators in \eqref{eq1} are a standard method in quantum information theory \cite{Sudarshan1961,Kraus1983,Nielsen2000,Jam2012,Chruscinski2014}, and they are not a part of the critique here. These operators were also used in the PRA article (see Eq. (4) in Ref.~\cite{Czerwinski2022}). However, the reasoning that follows these operators in the OE article contains two significant issues.

\subsubsection{Description of the absence of photon loss in the idler arm}

The authors of the OE article assume that no photon loss occurs in the idler arm, which they describe as follows:

\begin{center}
    {\it "\textcolor{red}{For the idler photon part, there is no photon loss, so it remains as $K^i = \ket{1}_i\!\bra{1}_i$.}"} \cite{Sengupta2024}
\end{center}

This description, found on page 40153 of the OE article, is inaccurate because it fails to clearly distinguish between the Kraus operators, which act on quantum states, and the quantum states themselves. If there is no photon loss in the idler channel, the Fock state can be represented as $\varrho_i = \ket{1}_i\!\bra{1}_i$. While this Fock state remains constant, the notation used by the authors is ambiguous and suggests they intended to reference the Kraus operator. If one wishes to describe the transformation of the Fock state of the idler photon using Kraus operators, the correct approach would involve the identity operator, which leaves the photon-number state unchanged. That is, we could write $K^i = \mathbb{I}_2$. This operator can also be obtained by substituting $\eta = 1$ into \eqref{eq1}, which provides a trivial quantum operation with a unitary operator: $\Lambda_{1} [\varrho_i] = \mathbb{I}_2 \varrho_i \mathbb{I}_2 = \varrho_i.$

\subsubsection{Alleged two approaches to photon loss modeling}

The authors of the OE article claim that there are supposedly two approaches to describing lossy quantum channels. According to them, one approach is to use the Kraus operators as in \eqref{eq1} and the other involves an operator $M_1^s$ given in Eq. (7) of the OE article. However, we strongly disagree with this interpretation as we believe Eq. (7) of the OE article was misinterpreted by again confusing the Kraus operator with the quantum state. To clarify, let us restate Eq. (7) and mark it in red to indicate a misconception:
\begin{equation}\label{nibykraus}
 \color{red}{M_1^s  = (1- \eta) \ket{0}_s \!\bra{0}_s +  \eta \ket{1}_s \!\bra{1}_s}\tag{7}.
 \end{equation}
This structure is misidentified as a {\it Kraus operator}, when in fact it does not meet the criteria for such an operator. The representation above actually results from the action of the Kraus operators in \eqref{eq1} on the initial Fock state in the signal channel, i.e., $\varrho_s = \ket{1}_s \!\bra{1}_s$, which can be expressed as
\begin{equation}\label{eq2}
    \Lambda_{\eta} [\varrho_s] = K_0^s  \varrho_s (K_0^s)^{\dagger} + K_1^s  \varrho_s (K_1^s)^{\dagger} =  (1- \eta) \ket{0}_s \!\bra{0}_s +  \eta \ket{1}_s \!\bra{1}_s. \tag{C1}
\end{equation}

Consequently, Eq. (7) of the OE article should be interpreted as the quantum state representing the photon number after propagation through a medium with transmittance $\eta$. However, the authors of the OE article falsely treat it as a Kraus operator and state on page 40154 as follows
\begin{center}
    {\it "\textcolor{red}{Finally, the total Kraus operator becomes}" \cite{Sengupta2024}}
    \begin{equation}\label{eqkraus}
     \textcolor{red}{M = M_1^s \otimes M_1^i = \left( (1- \eta) \ket{0}_s \!\bra{0}_s +  \eta \ket{1}_s \!\bra{1}_s \right)\otimes \ket{1}_i\!\bra{1}_i } \tag{C2}
    \end{equation}
\end{center}

The object denoted by $M$ cannot be considered a Kraus operator. As explained, the authors of the OE article incorrectly treat Fock states as Kraus operators. In the formula above, $M_1^s$ corresponds to the Fock state in the signal arm \eqref{eq2}, and $M_1^i$ to that in the idler arm. Neither is a Kraus operator, and therefore they cannot describe quantum operations in a lossy channel.

In addition, the alleged second approach to photon loss modeling can be disproved by recognizing the fact that photon loss is inherently a non-unitary process \cite{Czerwinski2022}. Consequently, it requires more than a single Kraus operator for a complete and accurate description. Therefore, a single operator denoted by $M$ in \eqref{eqkraus} cannot possibly account for this phenomenon.

\subsection{Polarization entanglement under lossy channels}

The authors of the OE article consider a composite quantum state consisting of both a polarization-encoded state, representing an entangled pair of photons, and a Fock state, which represents the photon count in the corresponding arms (see Eq. (5) in Ref.~\cite{Sengupta2024}). However, they assume that the polarization-entangled state remains unaffected by photon loss, modeling it by an identity operator $\mathbb{I}_2 \otimes \mathbb{I}_2$ acting on this part of the composite system (see Eq. (8) in Ref.~\cite{Sengupta2024}). Below Eq. (8) on page 40154 they state that 

\begin{center}
{\it "\textcolor{red}{The operation $\mathbb{I}_2 \otimes \mathbb{I}_2$ indicates that the photon loss does not change the polarization information.}"} \cite{Sengupta2024}
\end{center}

This approach is fundamentally flawed. Photon loss in one or both arms of an entangled photon pair inherently disrupts the entanglement, as it leads to the absence of a corresponding photon in the partner arm, thus breaking the correlation that defines the entangled state. In other words, photon loss not only reduces the number of photons in the system but also inevitably impacts the polarization state, leaving the remaining single photon (if any) in a mixed, non-entangled state. The unaffected application of an identity operator on the polarization part, as assumed in the OE article, disregards this essential effect of photon loss on the entangled state.

In reality, photon loss in one arm necessitates a corresponding transformation in the polarization-encoded state to reflect the transition from an entangled state to a mixed state, as detailed in Sec.~\ref{section2}. Ignoring this effect fundamentally misrepresents the behavior of quantum systems under lossy conditions and results in a misleading interpretation of the entangled state after propagation through a lossy channel.

\subsection{Density matrix of a composite state under lossy channels}

In Eq. (8) of the OE article, the authors propose an operator given by $\mathbb{I}_2 \otimes \mathbb{I}_2 \otimes M$, where the first two identity operators act on the polarization-encoded state of an entangled photon pair, while the operator $M$ is intended to describe photon loss in the Fock state. To illustrate the problem, let us recapture the equation
\begin{equation}\label{equation8}
  \textcolor{red}{\rho(\eta) = (\mathbb{I}_2 \otimes \mathbb{I}_2 \otimes M) \rho_{int} (\mathbb{I}_2 \otimes \mathbb{I}_2 \otimes M)^{\dagger}}. \tag{8}
\end{equation}

This formulation of the quantum channel in Eq. (8) of the OE article is fundamentally incorrect and cannot accurately represent the density matrix of a composite quantum state under photon loss. Specifically, $M$ is the same operator as defined in \eqref{eqkraus}. As discussed in prior subsections, the authors repeatedly confuse Kraus operators, which model quantum operations, with quantum states themselves. Here, $M$ is not a Kraus operator but rather a product of two Fock states, one for the signal (subject to photon loss) and one for the idler (unchanged). A correct description of photon loss requires a set of Kraus operators that account for the probabilistic nature of loss events in both arms, which would then act on the composite state. However, in \eqref{equation8}, $M$ acting on the photon-number state is in fact quantum state acting on a quantum state, which is not a correct description of this physical process.

Moreover, the authors model the entire process with a single operator, implicitly suggesting that photon loss can be described by a unitary transformation. This assumption is physically inaccurate: photon loss is inherently a non-unitary process, as it entails an irreversible change in the quantum system by breaking entanglement and leading to decoherence in the polarization state. The operator $\mathbb{I}_2 \otimes \mathbb{I}_2 \otimes M$ thus fails to capture the non-unitary nature of photon loss and the associated transformation from an entangled to a mixed state, as explained in Sec.~\ref{section2}.

In conclusion, the approach in Eq. (8) of the OE article misrepresents the effects of a lossy channel on both the Fock state and the polarization-entangled state. Proper modeling of such a composite system under lossy conditions requires a formalism based on Kraus operators that accurately represents the probabilistic and non-unitary effects of photon loss on the entire composite quantum state.

\subsection{Faulty method for analyzing quantum correlations after photon loss}

After introducing Eq. (8), the OE article proposes a method for evaluating quantum correlations in a lossy channel by tracing out the photon-number degrees of freedom and subsequently analyzing the entanglement properties of the remaining state. Again, let us cite a specific equation from the OE article:
\begin{equation}
    \textcolor{red}{\rho_{red} = \mathrm{Tr}_{N_1, N_2} \left( \rho(\eta) \right)}  \tag{9}.
\end{equation}

The authors calculate the reduced density matrix $\rho_{red}$ by tracing out the photon-number information and claim to measure quantum correlations through the CHSH inequality. They describe this process as follows on page 40154:
\begin{center}
    {\it \textcolor{red}{"In the first case discussed Eq. (6), the CHSH value remains constant and maximum even when the object’s reflectivity is very small. This helps us to confirm the presence of the object but does not assist in estimating the value of $\eta$ for the object."} } \cite{Sengupta2024}
\end{center}

The above statement is a natural consequence of the faulty assumptions underlying their model. Since the authors assumed that the polarization state remains unaffected by photon loss in one of the channels, tracing out the photon-number degrees of freedom results in what appears to be a maximally entangled state. However, this assumption is fundamentally incorrect, as it would imply that, even for very low transmittance (where photons in one arm are almost completely lost), one could still detect quantum entanglement using single-photon counts from the other arm. This approach is flawed, as the act of tracing out the Fock state cannot reverse the physical effects of photon loss on the polarization-encoded state. Photon loss inherently leads to decoherence, which disrupts the initial entanglement. Consequently, tracing out the photon-number information does not restore the entanglement in the remaining state; instead, the model used in the OE article obscures the fact that photon loss has already broken the correlations. As we describe in Sec. \ref{section2}, the correct method must account for how photon loss impacts the polarization-encoded state. By omitting this step, the authors’ approach leads to an artificially high CHSH value, as their reduced density matrix does not accurately represent the physical state after photon loss.

The authors further claim:
\begin{center}
    {\it \textcolor{red}{"However, in the second case Eq. (7), as the photon loss in the signal path increases, we record a decrease in the CHSH value, which helps to estimate the reflectivity of the object."} } \cite{Sengupta2024}
\end{center}

This statement also lacks justification in light of the present analysis. As explained previously, the object \eqref{nibykraus} is not a Kraus operator, and therefore any analysis regarding the impact of this joint operator on the initial composite state, as given in \eqref{equation8}, cannot lead to reliable results. The transformation used by the authors of the OE article is unphysical, making all results derived from this model highly questionable.

\subsection{Kraus operators for a depolarizing channel}

The lack of a fundamental understanding of the Kraus representation is once again evident on page 40156 of the OE article, where the authors attempt to represent a depolarizing channel using Kraus operators. In Eq. (11) of Ref.~\cite{Sengupta2024}, the density matrix after the action of a depolarizing channel is expressed as
\begin{equation}\label{eq11}
    \textcolor{red}{\rho (p) =  \sum_{i=0}^3 (K(p)_i)^{\dagger} \rho (K(p)_i) }\tag{11}.
\end{equation}

A similar expression is also provided in Eq. (12) of the OE article. However, we assert that both Eq. (11) and Eq. (12) are fundamentally incorrect. To clarify, we refer to the correct definition of a legitimate quantum channel acting in the operator space \( B(\mathcal{H}) \) associated with the Hilbert space \( \mathcal{H} \), as discussed in Ref.~\cite{CzerwinskiSym}.

\begin{definition}[\textbf{Quantum channel}]
A linear map $\Lambda:\: B(\mathcal{H}) \rightarrow B(\mathcal{H})$ is a quantum channel if and only if:
\begin{equation}\label{eqC3}
 \forall \hspace{0.25cm} X \in B(\mathcal{H}) \hspace{0.5cm} \Lambda[X] = \sum_{\alpha} K_{\alpha} \,X \, K_{\alpha}^{\dagger} \tag{C3}
\end{equation}
and
\begin{equation}\label{eqC4}
\sum_{\alpha} K_{\alpha}^{\dagger}\, K_{\alpha} = \mathbb{I}_d. \tag{C4}
\end{equation}
\end{definition}

To be a legitimate quantum channel, a map must be both completely positive (CP) and trace-preserving (TP); such maps are commonly referred to as CPTP maps. Complete positivity is guaranteed by the Kraus representation given in \eqref{eqC3}, while the condition in \eqref{eqC4} ensures that the map $\Lambda$ is trace-preserving \cite{Alicki1987,Bengtsson2006}. Common examples of quantum channels within quantum information theory include the bit flip, phase flip, depolarizing, amplitude damping, and phase damping channels, as described in Ref.~\cite{Nielsen2000}.

When comparing Eq. (11) and Eq. (12) of the OE article with the actual Kraus representation \eqref{eqC3}, it is clear that the authors have incorrectly placed the conjugate transpose symbol. Although one might argue that for self-adjoint (Hermitian) operators this mistake does not alter the outcome, we emphasize the importance of mathematical rigor in the operator representation of quantum channels. The proper way to represent a CPTP map is to adhere to the structure provided in \eqref{eqC3} and \eqref{eqC4}.

\vspace{0.75cm}
In conclusion, this section has demonstrated that the mathematical model used in the OE article is incorrect, as the operators do not describe actual physical processes. Moreover, the treatment of entanglement as unaffected by photon loss contradicts established knowledge on the nature of quantum entanglement. For this reason, this commentary will not go into the details of specific results presented in the article, as we believe these findings cannot be rationalized within the context of the model used. In other words, by disproving the foundational assumptions of the OE article’s model, we show that the specific findings are also questionable, as they do not arise from physically reliable assumptions.

In particular, the primary claim of the OE paper is the purported ability to confirm the presence of an object with very low reflectivity by utilizing quantum correlations. However, we argue that this claim lacks evidence. The authors mistakenly assumed that polarization entanglement remains unaffected by the lossy channel, leading to their conclusion that a violation of the CHSH inequality could be recorded despite almost complete loss of signal in one arm. Based on our analysis, we argue that this approach and its conclusions are unjustified.

\section{Mathematical Framework of Photon Loss For Polarization-entangled Photon Pairs}\label{section2}

In this section, we present a detailed description of the impact of photon loss on polarization-entangled photon pairs. This analysis is divided into two subsections. First, in Sec.~3.1., we examine the effect of photon loss on the quantum state encoded in the polarization degree of freedom. Then, in Sec.~3.2., we investigate how photon loss affects the photon-number state, represented by Fock states, with regard to photon detection probabilities.

\subsection{Impact of photon loss on the quantum state encoded in the polarization degree of freedom}\label{polarization}

When dealing with polarization-entangled photon pairs, photon loss (often caused by absorption or scattering during transmission) has a significant impact on the entangled state. Suppose we start with an entangled Bell state for two photons, described by the density matrix $\rho_{AB}$, given by:
\begin{equation}
    \rho_{AB} = \ket{\Phi^+}\!\bra{\Phi^+} = \frac{1}{2} \left( \ket{HH} + \ket{VV} \right) \left( \bra{HH} + \bra{VV} \right),\tag{C5}
\end{equation}
where $\ket{H}$ and $\ket{V}$ represent the horizontal and vertical polarization states, respectively. This state features maximal entanglement in the polarization degree of freedom since it belongs to the class of Bell states.

When one of the photons, say photon \( B \), is lost due to attenuation, we effectively trace out the degrees of freedom associated with \( B \). Mathematically, this is represented by taking the partial trace over the subsystem \( B \):
\begin{equation}
    \rho_A = \mathrm{Tr}_B (\rho_{AB}).\tag{C6}
\end{equation}
Performing the partial trace on $\rho_{AB}$ yields a reduced density matrix for photon \( A \):
\begin{equation}
    \rho_A = \frac{1}{2} \left( \ket{H}\!\bra{H} + \ket{V}\!\bra{V} \right),\tag{C7}
\end{equation}
which represents a maximally mixed state, i.e. $\rho_A \equiv 1/2 \:\mathbb{I}_2$. This outcome indicates that the quantum correlations of the entangled state has been lost, as photon \( A \) alone no longer retains any information about the original entanglement.

The loss of entanglement can be interpreted as a transition from a pure, entangled state to a probabilistic mixture, where the state of the surviving photon (photon \( A \)) is independent of any specific polarization outcome. The reduced density matrix $\rho_A$ is diagonal, signifying a complete loss of quantum coherence in the polarization basis, thus marking the absence of entanglement.

In general, if the entanglement is represented by a pure density matrix $\rho_{AB}$, photon loss due to attenuation transforms it into a mixed state $\rho_A$ for the remaining photon. Consequently, photon loss in one arm of an entangled photon pair leads to the degradation of entanglement, leaving the unaffected photon in a mixed state that reflects no residual entanglement.

\subsection{Impact of photon loss on the Fock state representing the number of photons}\label{fockstate}

To further investigate photon loss for polarization-entangled pairs, we can consider the photon-number states, or Fock states, which represent specific photon counts in each arm. Let us denote the transmittance in arms \( A \) and \( B \) by \( T_A \) and \( T_B \), respectively. In quantum mechanics, the transmittance \( T_X \) corresponds to the probability of detecting a photon at the end of arm \( X \), while \( 1 - T_X \) denotes the probability of photon loss in that arm. To describe the photon counts for an entangled pair, we have to implement the joint probability distribution across both arms. Thus, the probability of detecting one photon in each arm after transmission is given by the product \( T_A T_B \). Similarly, we can determine the probabilities for all other possible detection events.

We begin with the initial Fock state \( \ket{1}_A \ket{1}_B \), indicating one photon in each arm. Introducing loss in one or both arms modifies the photon-number distribution, resulting in a mixed state composed of four possible detection outcomes. After accounting for transmission probabilities, the state transforms into a mixture of the following possibilities:
\begin{itemize}
    \item \textbf{Case 1: One photon detected in each arm; probability \( T_A T_B \)}.\\
    With probability \( T_A T_B \), both photons are successfully transmitted, resulting in the final state \( \ket{1}_A \ket{1}_B \).
    \item \textbf{Case 2: Zero photons in arm \( A \), one photon in arm \( B \); probability: \( (1 - T_A) T_B \)}.\\
    Photon loss in arm \( A \) occurs with probability \( 1 - T_A \) while the photon in arm \( B \) is successfully transmitted. This leads to the state \( \ket{0}_A \ket{1}_B \).
    \item \textbf{Case 3: One photon in arm \( A \), zero photons in arm \( B \); probability: \( T_A (1 - T_B) \)}.\\
    Here, the photon in arm \( A \) is transmitted with probability \( T_A \), while the photon in arm \( B \) is lost with probability \( 1 - T_B \), resulting in the state \( \ket{1}_A \ket{0}_B \).
    \item \textbf{Case 4: Zero photons in both arms; probability: \( (1 - T_A)(1 - T_B) \)}.\\
    With probability \( (1 - T_A)(1 - T_B) \), both photons are lost, leading to the vacuum state \( \ket{0}_A \ket{0}_B \).
\end{itemize}

The photon-number state after transmission, therefore, becomes a mixed state composed of these four cases:
\begin{equation}\label{eqC6}
\begin{split}
     \varrho_{\text{out}} = & T_A T_B \ket{1}_A \ket{1}_B \!\bra{1}_A \bra{1}_B + (1 - T_A) T_B \ket{0}_A \ket{1}_B \!\bra{0}_A \bra{1}_B + T_A (1 - T_B) \ket{1}_A \ket{0}_B \! \bra{1}_A \bra{0}_B \\&+ (1 - T_A)(1 - T_B) \ket{0}_A \ket{0}_B \! \bra{0}_A \bra{0}_B.
\end{split}\tag{C8}
\end{equation}

This density matrix, \( \varrho_{\text{out}} \), represents a probabilistic mixture of all possible photon detection outcomes after photon loss in both arms. Each term corresponds to one of the four detection scenarios, weighted by its respective probability. The resulting state captures the effects of photon loss on the Fock state representation and illustrates how the original entangled state degrades into a mixture of different photon-number configurations. This formulation is useful for understanding photon loss in quantum communication channels, where the probability of detecting photons in each arm can significantly impact the performance of quantum protocols relying on entanglement.

In the quantum illumination model presented in the OE article, the authors assume that only one channel is lossy due to the presence of an object with a reflectivity $\eta$. While this assumption captures the core concept of quantum illumination, it appears that treating the other channel as lossless is an oversimplification. Due to their inherent fragility, photons are inevitably subject to losses in both channels, making a zero-loss channel an overly optimistic scenario. Here, we present a more general model that treats both channels as imperfect, resulting in a more comprehensive decomposition of the two-photon Fock state, as shown in \eqref{eqC6}.

\section{Unreferenced Utilization of Prior Work on Photon Attenuation}\label{section1}

Unfortunately, we must point out the glaring lack of attribution to Ref. \cite{Czerwinski2022} by the authors of Ref. \cite{Sengupta2024}. Although the PRA article appears in the bibliography and is cited in the last paragraph of the introduction, it is later used in a completely different context. In Sec. 3.3, titled "Attenuation and range estimation," the OE article presents equations directly taken from Ref. \cite{Czerwinski2022} without proper reference. This is a significant issue, as the general mention of Ref. \cite{Czerwinski2022} in the introduction is not a sufficient attribution. The equations in Sec. 3.3 of the OE article are presented in such a way that they appear as if they were original contributions of the authors, which misleads the reader.

In Table \ref{tab:eqcompare}, we present equations that appear in both publications and are borrowed in Ref. \cite{Sengupta2024} without citation. The symbols in these equations are identical, suggesting that this repetition is not coincidental.

\begin{table}[h!]
\centering
\renewcommand{\arraystretch}{2.0}
\begin{tabular}{|c|c|c|}
\hline
\textbf{Equation} & \textbf{Ref. \cite{Czerwinski2022}} (the PRA article) & \textbf{Ref. \cite{Sengupta2024}} (the OE article) \\ \hline
$\rho(L) = \sum_{j=0}^{N} \binom{N}{j} \left( e^{-\Lambda L} \right)^j \left( 1 - e^{-\Lambda L} \right)^{N-j} | j \rangle \langle j |$ & Eq. (8) & Eq. (14) \\ \hline
$\Lambda \equiv \frac{\ln 10 \:\alpha}{10}$ & below Eq. (1) & below Eq. (14) \\ \hline
$\rho(L) = \left( 1 - e^{-\Lambda L} \right) |0\rangle \langle 0| + e^{-\Lambda L} |1\rangle \langle 1|$ & Eq. (2) & Eq. (16) \\ \hline
\end{tabular}
\caption{Table showing equations borrowed by the authors of Ref. \cite{Sengupta2024} from Ref. \cite{Czerwinski2022} without citations.}
\label{tab:eqcompare}
\end{table}

The authors of the OE article not only directly copied specific elements of the PRA article but also replicated the analytical approach originally introduced in Ref.~\cite{Czerwinski2022}. In the PRA article, the decoherence of the Fock state was quantified and analyzed by computing the purity and the von Neumann entropy (see Eqs. (13) and (14) in Ref.~\cite{Czerwinski2022}). While purity and entropy calculations are well-known tools, the specific use of these metrics to assess the Fock state’s behavior as propagation distance increases was an original insight of the PRA article, as demonstrated in Fig. 2 in Ref.~\cite{Czerwinski2022} and discussed in the corresponding analysis.

Similarly, in the OE article, we find an identical pair of figures of merit -- the purity and the von Neumann entropy in Eq. (15) -- used to examine the decoherence of the Fock state. Additionally, Fig. 7 in Ref. \cite{Sengupta2024} includes an inset plot conceptually identical to Fig. 2 in Ref. \cite{Czerwinski2022}, though with modified data to reflect a different attenuation coefficient.

In conclusion, a significant portion of Sec. 3.3 of the OE article relies on the framework developed in the PRA article, without proper acknowledgment. The absence of citations in Sec. 3.3 creates the impression that this analysis is an original contribution of the OE article authors, while it closely mirrors the approach introduced in the PRA article from 2022. Finally, it is worth noting that the model for decoherence of the Fock state described in the PRA article is appropriate for fiber-based transmission, where photon loss is primarily due to absorption and scattering, which can be represented by a simple damping factor. However, this model is not suitable for FSO transmission, where additional phenomena contribute to overall photon loss. We discuss these issues in detail in Sec.~\ref{section3}, where we enumerate several processes that affect the transmission of optical signals through the atmosphere.

\section{Photon Loss Modeling in FSO Transmission}\label{section3}

In Sec. 3.3., the OE article attempts to model photon loss in FSO transmission by implementing only the Beer-Lambert law with a fixed attenuation coefficient to represent the effect of atmospheric losses. The authors assume an attenuation coefficient of $0.07$ dB/km, which is significantly lower than the typical attenuation found in telecommunication fibers. While it is true that atmospheric attenuation due to absorption and scattering is less significant than in standard fibers, the OE model fails to capture other phenomena that contribute to the overall photon loss in FSO transmission. By treating the atmosphere as if it were an optical fiber with low attenuation, the authors neglect a variety of physical processes that affect photon transmission through the air. As a result, their approach is oversimplified, which implies that the findings and conclusions about photon loss and quantum correlations in free-space transmission are not reliable.

Free-space optical transmission through the atmosphere is a complex process that requires consideration of several distinct factors beyond simple attenuation. Key phenomena that should be included in a comprehensive model of FSO transmission include \cite{Andrews2005,Pirandola2021,Trinh2022}:

\begin{itemize}
  \item \textbf{Beam Broadening:} In free-space transmission, the beam undergoes inevitable broadening due to diffraction. As it propagates over a distance \( z \), the waist of the beam increases, leading to an expanded spot size. Consequently, only a fraction of the beam can be captured by the receiver's aperture, resulting in geometrical losses as portions of the broadened beam fall outside the receiver's range.

    \item \textbf{Atmospheric Turbulence:} Turbulence caused by temperature and pressure variations in the atmosphere can lead to fluctuations in the refractive index, distorting the beam's phase and amplitude. This causes beam wander, scintillation, and phase distortions, all of which impact the signal quality.

    \item \textbf{Pointing Errors:} In practical FSO systems, maintaining precise alignment between the transmitter and receiver is challenging, especially over long distances or when environmental factors such as wind and vibrations come into play. Misalignment leads to pointing errors, which can reduce the power received and further increase photon loss.

    \item \textbf{Weather Conditions:} Factors like rain, fog, and clouds can heavily attenuate the signal, as water droplets and other particles in the atmosphere increase scattering and absorption.

\end{itemize}

The minimal effect that a realistic FSO transmission model must include is the loss due to beam broadening, also referred to as geometrical loss. This loss is inevitable in FSO transmission because, unlike in fiber-based systems where the beam is confined, an FSO signal naturally diverges over distance, leading to a reduction in the received signal power.

To demonstrate the significance of geometrical loss in FSO transmission, we compare the model presented in the OE article with our expanded model, which incorporates this critical factor. This comparison highlights the limitations of the OE approach and demonstrates the necessity of including geometrical loss for a physically accurate representation of photon loss in FSO transmission.

In our model, we assume that the overall transmittance for a horizontal FSO channel of length \( z \) is a product of two factors:
\begin{equation}
    T_{atm} (z) = 10^{- \alpha z} \:\eta_{\textrm{geo}} (z),\tag{C9}
\end{equation}
where \( \alpha \) denotes the attenuation coefficient, and \( \eta_{\textrm{geo}} (z) \) represents the geometrical loss. Based on Ref.~\cite{Pirandola2021}, the diffraction-induced factor \( \eta_{\textrm{geo}} (z) \) can be expressed as
\begin{equation}
    \eta_{\textrm{geo}} (z) = 1 - e^{- 2 a^2_R/ w^2(z)},\tag{C10}
\end{equation}
where \( a_R \) denotes the radius of the receiving telescope and \( w(z) \) represents the increased spot size:
\begin{equation}\label{eqC9}
    w(z) = w_0 \sqrt{1 + \left(\frac{z}{z_R}\right)^2},\tag{C11}
\end{equation}
with \( z_R = \pi w_0^2 \lambda^{-1} \) denoting the Rayleigh range and \( w_0 \) as the initial spot size. For our simulations, we assume \( \lambda = 1550 \) nm and \( w_0 = 1 \) cm. Additionally, for efficient comparison, we use the attenuation coefficient from the OE article, i.e. \( \alpha = 0.07 \) dB/km.

\begin{figure}[h] {\includegraphics[width=0.7\columnwidth]{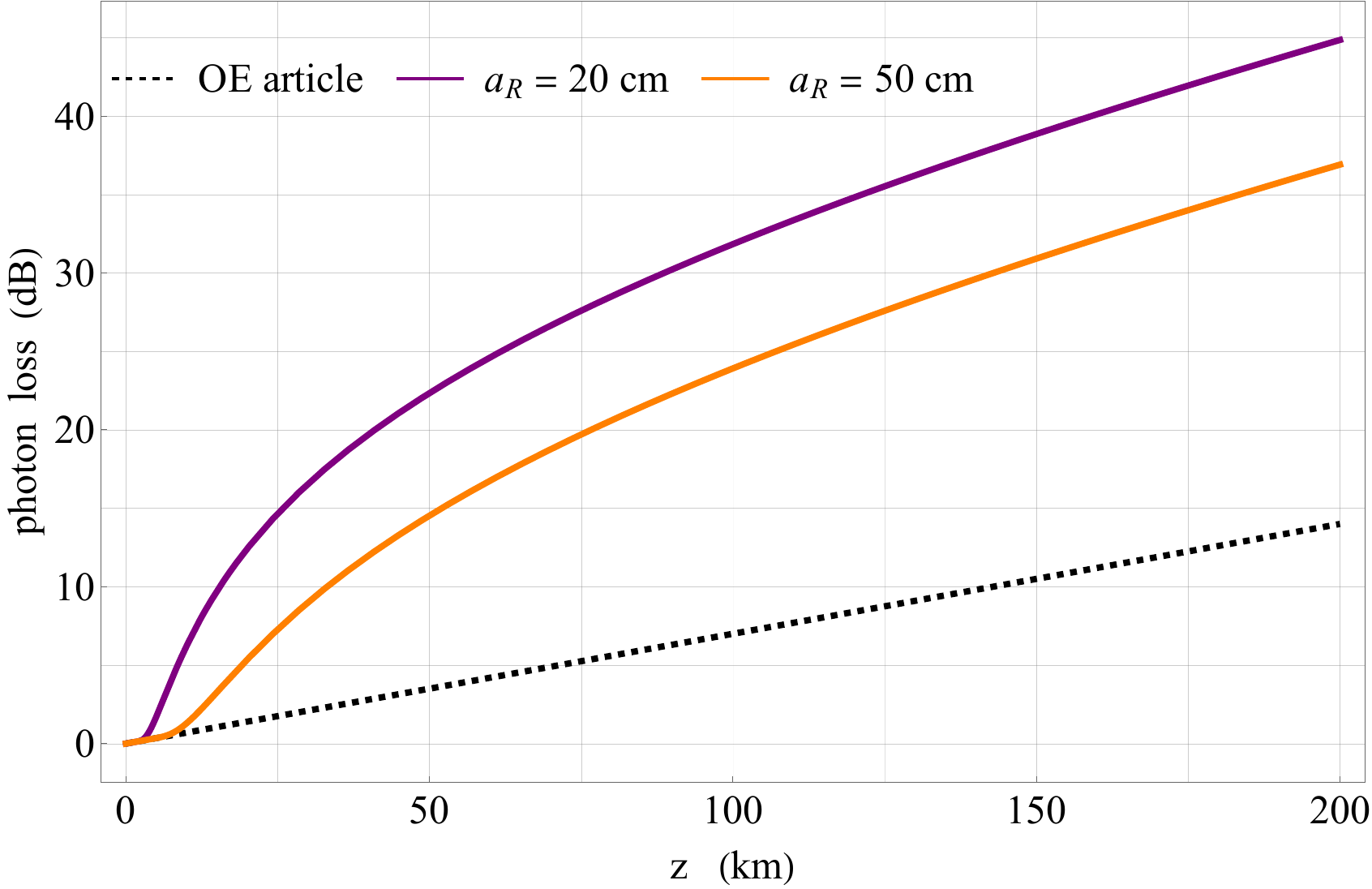}}
	\caption{Comparison of photon loss in FSO transmission according to the OE article (dashed line) and the improved model that accounts for geometrical loss.}
	\label{figure1}
\end{figure}

In \figref{figure1}, we present a comparison of photon loss (given in dB) versus the propagation distance. We evaluated two receiver telescope sizes: 20 cm and 50 cm. Even with a 50 cm radius, the losses are significantly greater than in the OE model. This demonstrates that the OE analysis relied on assumptions that were unrealistic, resulting in loss estimates that were excessively low.

As mentioned earlier, modeling the transmission of an optical signal through the atmosphere is complex and challenging. Our goal here is not to provide an exhaustive model but to show the shortcomings of the OE model by including one typical source of photon loss in FSO transmission. The results indicate that the OE model’s loss estimates are unreasonably low for realistic FSO transmission conditions.

\section{Discussion and Conclusions}\label{finalsec}

In this critique, we identified key errors in the OE article \cite{Sengupta2024} concerning the treatment of Kraus operators and the assumptions regarding photon loss in entangled systems. Specifically, the authors incorrectly identify quantum states as Kraus operators, leading to an inaccurate description of photon loss. Additionally, the OE article suggests two approaches to describing lossy quantum channels. However, only the first approach, which uses Kraus operators, correctly reflects the non-unitary nature of photon loss, while the second, involving a single operator, fundamentally misunderstands this phenomenon.

Furthermore, the OE article incorrectly assumes that photon loss does not impact the polarization-encoded entangled state, leading to flawed conclusions about the behavior of entanglement in a lossy channel. Photon loss inherently disrupts entanglement, transitioning the affected state to a mixed state and breaking polarization correlations between photons in different arms. Consequently, their approach, which treats photon loss as a process that does not affect polarization entanglement, misrepresents the behavior of quantum correlations subject to lossy channels. By tracing out photon-number degrees of freedom, they supposedly retain maximal entanglement, resulting in conclusions inconsistent with the real impact of lossy channels on quantum systems. Properly addressing photon loss requires a comprehensive Kraus operator approach, capturing the probabilistic and non-unitary effects of such loss on composite quantum states.

In addition, the criticized OE paper shares a strikingly similar analytical framework as the PRA article \cite{Czerwinski2022} for assessing photon attenuation in Fock states, specifically using the purity and the von Neumann entropy as metrics to quantify decoherence with distance. Originally introduced in the PRA article, the framework is applied identically in Sec. 3.3 of the OE article without citation, alongside comparable visual elements such as Fig. 7 in the OE article, which mirrors Fig. 2 in the PRA article. This substantial overlap suggests that key analytical insights and tools from the PRA article were reused in the OE article without proper attribution.

In this comment, we proposed a more accurate approach to address the impact of photon loss in both polarization-encoded and photon-number states, emphasizing how different types of quantum states respond to photon attenuation. For polarization-encoded photon pairs, we demonstrated that photon loss leads to a complete loss of entanglement, as the remaining photon transforms into a maximally mixed state without any quantum coherence in the polarization basis. In contrast, for photon-number states represented by Fock states, photon loss introduces a mixture of detection probabilities across both arms, depending on the transmittance values in each channel. This probabilistic mixture captures the degradation of the initial state as photon loss results in a combination of different photon-number configurations.

Additionally, the oversimplified approach to FSO transmission in the OE article, which relies solely on a fixed attenuation coefficient derived from the Beer-Lambert law, neglects additional loss factors inherent to the atmosphere, such as atmospheric turbulence, pointing errors, and varying weather conditions. In particular, our expanded model incorporates geometrical loss due to beam broadening. This improved modeling approach ensures a more accurate representation of photon loss, which can be used for reliable predictions of quantum correlations in FSO transmission.

In conclusion, while we do not question the validity of the experimental results presented in the OE paper, we have provided sufficient arguments to challenge the mathematical framework employed by the authors. Our analysis shows that the theoretical assumptions and formulas used in the OE article are flawed, which undermines the reliability of their findings. Consequently, there is no substantial evidence to support the authors' main conclusion of successfully detecting objects with very low reflectivity using the described methods. Although the experimental setup proposed in the OE paper could hold potential significance for quantum illumination, we argue that the theoretical framework of this research is based on incorrect assumptions, rendering all conclusions drawn from this investigation unreliable.


\begin{thebibliography}{1}

\bibitem{Sengupta2024}
K. Sengupta, K.M. Shafi, S. Asokan, and C.M. Chandrashekar, Quantum illumination using polarization-entangled photon pairs for enhanced object detection. \href{https://doi.org/10.1364/OE.531674}{Opt. Express} {\bf 32}, 40150-40164 (2024). Preprint version: \href{https://arxiv.org/html/2401.10182v2}{arXiv:2401.10182v2}


\bibitem{Czerwinski2022}
A. Czerwinski and J. Szlachetka, Efficiency of photonic state tomography affected by fiber attenuation. \href{https://doi.org/10.1103/PhysRevA.105.062437}{Phys. Rev. A} {\bf 105}, 062437 (2022).

\bibitem{Sudarshan1961}
E.C.G. Sudarshan, P.M. Mathews, and J. Rau, Stochastic Dynamics of Quantum-Mechanical Systems. \href{https://doi.org/10.1103/PhysRev.121.920}{Phys. Rev.} {\bf 121}, 920 (1961).

\bibitem{Kraus1983}
K. Kraus, States, {\it Effects, and Operations—Fundamental Notions
of Quantum Theory}, (Springer, Berlin, Germany, 1983).

\bibitem{Nielsen2000}
M.A. Nielsen and I.L. Chuang, {\it Quantum Computation and Quantum Information}, (Cambridge University Press, Cambridge, 2000).

\bibitem{Jam2012}
A. Jamiołkowski, {\it Fusion Frames and Dynamics of Open Quantum Systems}, [in:] S. Lyagushyn (Ed.), {\it Quantum Optics and Laser Experiments}, (InTech, Rijeka, Croatia, 2012); pp. 67–84.

\bibitem{Chruscinski2014}
D. Chruściński, On Time-Local Generators of Quantum Evolution. \href{https://doi.org/10.1142/S1230161214400046}{Open Syst. Inf. Dyn.} {\bf 21}, 1440004 (2014).

\bibitem{CzerwinskiSym}
A. Czerwinski, Dynamics of Open Quantum Systems—Markovian Semigroups and Beyond. \href{https://doi.org/10.3390/sym14081752}{Symmetry} {\bf 14}, 1752 (2022).

\bibitem{Alicki1987}
R. Alicki and K. Lendi, {\it Quantum Dynamical Semigroups and Applications}, (Springer, Berlin, 1987).

\bibitem{Bengtsson2006}
I. Bengtsson and K. Życzkowski, {\it Geometry of Quantum States}, (Cambridge University Press, New York,
2006).

\bibitem{Andrews2005}
L.C. Andrews and R.L. Phillips, {\it Laser beam propagation through random media. Second Edition}, (SPIE Press, Bellingham, Washington USA, 2005).

\bibitem{Pirandola2021}
S. Pirandola, Satellite quantum communications: Fundamental bounds and practical security. \href{https://doi.org/10.1103/PhysRevResearch.3.023130}{Phys. Rev. Research} {\bf 3}, 023130 (2021).

\bibitem{Trinh2022}
P.V Trinh, A. Carrasco-Casado, H. Takenaka, M. Fujiwara, M. Kitamura, M. Sasaki, and M. Toyoshima, Statistical verifications and deep-learning predictions for satellite-to-ground quantum atmospheric channels. \href{https://doi.org/10.1038/s42005-022-01002-1}{Commun. Phys.} {\bf 5}, 225 (2022).

\end{thebibliography}
\end{document}